%% file: main.tex
\documentclass[sigconf]{acmart}

\usepackage{fancyhdr}
\usepackage{lipsum} 

\PassOptionsToPackage{hyphens}{url}
\usepackage{hyperref}
\usepackage{multirow}
\usepackage{graphics}
\usepackage{enumitem}
\usepackage{color, colortbl}
\usepackage{lipsum}
\usepackage{tabularx,booktabs}
\usepackage[section]{placeins}
\usepackage{graphicx}
\usepackage{comment}
\usepackage{subcaption}
\usepackage[para]{footmisc}
\usepackage{rotating}
\usepackage{tablefootnote}
\restylefloat{table}
\definecolor{lightgray}{rgb}{0.83, 0.83, 0.83}
\usepackage{caption}
\usepackage{subcaption}

\DeclareCaptionLabelFormat{simplab}{Figure~#2}
\DeclareCaptionFormat{nocap}{}
\DeclareCaptionSubType*{figure}

\usepackage{caption}
\copyrightyear{2024}
\acmYear{2024}
\setcopyright{acmlicensed}\acmConference[CODASPY '24]{Proceedings of the Fourteenth ACM Conference on Data and Application Security and Privacy}{June 19--21, 2024}{Porto, Portugal}
\acmBooktitle{Proceedings of the Fourteenth ACM Conference on Data and Application Security and Privacy (CODASPY '24), June 19--21, 2024, Porto, Portugal}
\acmDOI{10.1145/3626232.3653282}
\acmISBN{979-8-4007-0421-5/24/06}

\begin{document}



\title{Smishing Dataset I: Phishing SMS Dataset from Smishtank.com \\
\large Data/Toolset paper}

\author{Daniel Timko}
\affiliation{%
  \institution{California State University San Marcos}
   \city{San Marcos}
   \state{CA}
   \country{USA}}
\email{timko002@csusm.edu}

\author{Muhammad Lutfor Rahman}
\affiliation{%
  \institution{California State University San Marcos}
   \city{San Marcos}
   \state{CA}
   \country{USA}}
\email{mlrahman@csusm.edu}

\begin{abstract}
\input{abstract.tex}
\end{abstract}

\maketitle

\input{introduction}

\input{previouswork}
\input{messageprocessing}
\input{results}
\input{discussion}
\input{conclusion}
\input{acknowledgement}

\bibliographystyle{plain}
\bibliography{main}

\end{document}

%% file: abstract.tex
\noindent While smishing (SMS Phishing) attacks have risen to become one of the most common types of social engineering attacks, there is a lack of relevant smishing datasets. One of the biggest challenges in the domain of smishing prevention is the availability of fresh smishing datasets. Additionally, as time persists, smishing campaigns are shut down and the crucial information related to the attack are lost. With the changing nature of smishing attacks, a consistent flow of new smishing examples is needed by both researchers and engineers to create effective defenses.
In this paper, we present the community-sourced smishing datasets from the smishtank.com. It provides a wealth of information relevant to combating smishing attacks through the breakdown and analysis of smishing samples at the point of submission. In the contribution of our work, we provide a corpus of 1062 smishing samples that have been publicly submitted through the site. Each message includes information relating to the sender, message body, and any brands referenced in the message. Additionally, when a URL is found, we provide additional information on the domain, VirusTotal results, and a characterization of the URL. Through the open access of fresh smishing data, we empower academia and industries to create robust defenses against this evolving threat. 

%% file: introduction.tex
\section{Introduction}

\hspace{0.4cm} Smishing, a variant of phishing that leverages text messages, has experienced a notable surge in frequency over recent years~\cite{FCCRobotext}. Responding to this escalating threat, the Federal Communications Commission (FCC) took an unprecedented measure by instructing mobile providers to proactively block messages that are deemed \textit{highly likely to be illegal}~\cite{FCCRulesChange}.  Prior research on smishing attacks indicates that victims often exhibit a high susceptibility, frequently engaging with these campaigns by clicking on links, responding to the texts, or calling numbers included in the messages~\cite{Raman2023, timkounveil23}. What makes phishing, and in particular smishing, attacks so difficult to prevent is their ever changing nature. Year over year, smishing trends and strategies change, introducing novel smishing attacks. Examples of this trend include a significant rise in COVID-related smishing incidents during the peak periods of the pandemic~\cite{irswarn}. 

To counter this growing threat, creating new publicly available datasets becomes imperative to develop effective models and strategies. However, there are limited publicly available SMS phishing datasets~\cite{Saeki2022,Mishra2021DSmishSMSAST}. These datasets also face relevance challenges due to various factors. As these messages age, linked websites are often taken down swiftly, limiting learning opportunities about the smishing attack~\cite{DatingPhish}. Moreover, older datasets become less effective as smishing campaigns evolve, leading to concept drift~\cite{Profiler2022}. Another issue is that some researchers consider smishing to be a subset of spam and group them together in datasets~\cite{8965429,10139070, doi:10.1080/19393555.2015.1078017}.
While phishing is inherently malicious, spam can encompass unsolicited advertising.

To enhance dataset robustness, it's crucial to collect pertinent information close to the attack time. Timko et al.~\cite{timko2023commercial} developed a framework for gathering and storing screenshots of SMS messages and released a dataset of 55 smishing samples. However, our work builds upon this concept to provide a larger dataset with a more comprehensive analysis of the messages. We also automated the process to parse message attributes and gather relevant WHOIS and VirusTotal information during submission. Furthermore, we manually categorized messages and URLs based on content, facilitating easier access to researchers without necessitating crawling other public sources.

Furthermore, the submitted messages should be parsed and analyzed for important information that may be useful in spam and phishing classification tasks. According to proofpoint's state of the phishing report~\cite{proofpointsotf}, brand abuse is an integral part of social engineering attacks. Therefore, any mentions of brands in the message contents should include them as an attribute, potentially key to identifying who the phisher is attempting to impersonate. Additionally, examining the WHOIS and VirusTotal information for URLs contained in a message can provide a historical perspective on domain history and security vendor ratings. Finally, several works have investigated the categorization of phishing URLs and SMS messages~\cite{cluesintwitter,bethephisher, insidephishersmind}. By using these categorizations we can gain insights into the types of messages used in modern phishing attacks and how they seasonally evolve.

As part of this approach, we present a dataset comprising 1062 phishing SMSes sourced from smishtank.com. smishtank.com aims to address the scarcity of relevant public data by continually updating fresh SMS phishing messages. Leveraging open-source and public API resources, Smishtank consolidates varied sources into a single accessible platform. Presently, no other websites offer users the opportunity to publicly submit smishing messages for collaborative dataset creation.

This paper makes the following contributions:

\begin{enumerate}[leftmargin=*]
   \item We make publicly available a corpus of 1062 smishing messages.
   \item In alternative to other approaches, our dataset is based on community submissions that are already categorized and parsed to generate relevant data fields. 
   \item Message metadata were collected at the time of submission, including VirusTotal and Domain WHOIS information. In result, we enable researchers and developers to better understand the messages at the point they were fresh.
 \end{enumerate}

%% file: previouswork.tex
\section{Related Works}
\label{sec:relatedworks}
\paragraph{The Need for New Smishing Data}
\hspace{0.4cm} Work by Salman et al.~\cite{salman2022empirical} highlights the availability of data as one of the most significant challenges in SMS filtering. In order to tackle the changing nature of phishing messages, fresh relevant data is needed for phishing corpus. Several approaches in recent years have been explored to collect these fresh SMS messages. Tang et al.~\cite{cluesintwitter} explored crawling the data from publicly reported SMS messages on Twitter. However, recent updates to X(Formally Twitter) strictly forbid data crawling. Reaves et al.~\cite{reaves2018} collected the largest public SMS corpus of over nine hundred thousand messages collected through sms gateways. While this message set is large, few messages resulted in smishing attacks. Additionally, these public gateways drop MMS messages and lack some important aspects of smishing, such as the sender~\cite{nahapetyansms}. 

\paragraph{Community-based Solutions for Research Data}
\hspace{0.4cm} In a related area of phishing, community sourced solutions for phishing URL datasets have been successful. Publicly available phishing URL feeds, including phishtank~\cite{phishtank} and the APWG~\cite{apwg}, provide phishing URL feeds that are often used in datasets for the community and researchers. These resources use community-based submissions which are then analyzed and voted on by the public. Additionally, these sources parse the message brands from websites being linked in the URLs. Aligned with these directions, our work intends to breach this gap for phishing SMS.

\paragraph{Message and URL Characterization}
\hspace{0.4cm} In the phishing and spam domain, several prior works have explored characterizing phishing URLs and messages. Work by Quinkert et al.~\cite{bethephisher}, discusses 5 different types of squatting techniques that are used by phishers to create domains similar to the authentic ones used by businesses. Several prior works have explored classifying sms spam messages into categories~\cite{cluesintwitter,reaves2018}. Oest et al.~\cite{Oest2018} analyzed phishing kits and created a high level classification of phishing URLs. By being able to identify these categories, you can understand aspects of how a phisher attempts to deceive the target of phishing campaigns.

%% file: messageprocessing.tex
\section{Message Collection \& Processing}
\label{sec:processing}
\begin{figure*}[h]
    \centering
    \includegraphics[width=\textwidth]{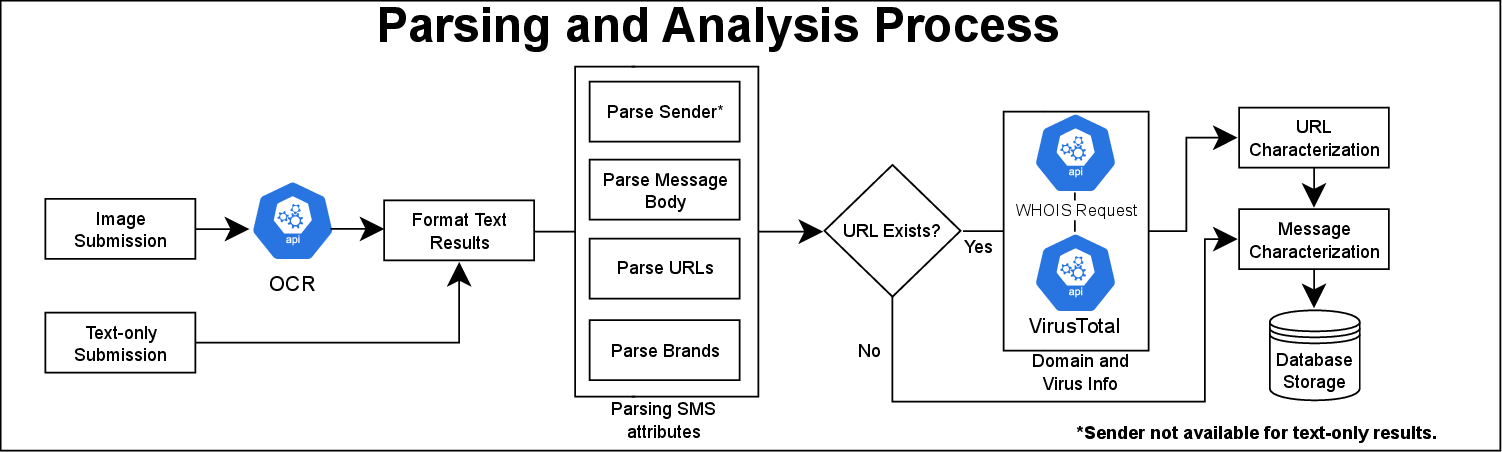}
    \caption{Message collection and processing procedure.}
    \label{fig:methodology}
    \vspace{-0.5cm}
\end{figure*}

\hspace{0.4cm} In the following sections we will describe the steps that the messages go through in the message processing phase. Next, we describe the features and analysis that we perform to gather additional information from messages submitted. The process for collecting, parsing and analyzing messages can be found in Figure~\ref{fig:methodology}. This research was conducted with the approval of our university's Institutional Review Board (IRB).

\subsection{Message Collection} 
\hspace{0.4cm} Messages were gathered exclusively from the smishing submission section on smishtank.com. These submissions comprise either a screenshot of the message or its text copy. Community-sourced datasets have become a prevalent and valuable resource in phishing research. In the realm of phishing URLs, Phishtank.com~\cite{phishtank} stands as an essential repository for new phishing URLs contributed by users. To augment submissions on the site, we promoted its features across social media platforms, encouraging users to share instances of smishing messages they received. Currently, we've amassed a total of 1062 validated submissions since the site's launch. Upon submission, these screenshots are stored in a data lake in a standard image format, with the extracted data incorporated into the dataset.

\subsection{Image to Text OCR}
\hspace{0.4cm}
To extract text from the SMS screenshots, we employed Optical Character Recognition (OCR) technology. This service converts the image into a readable text format, encompassing letters, words, lines, and paragraphs from the image's text content. Additionally, the OCR image recognition feature annotates the text's position within the image using rectangular bounding boxes. These boxes are defined by four sets of xy coordinates representing the corners of the rectangle enclosing the text. These bounding boxes provide contextual information for the text and aid in accurately identifying sections of the message. Moreover, each recognized piece of text is accompanied by a confidence interval, indicating the OCR's level of confidence in the accuracy of the identified text.

\subsection{Message Sender Parsing}
\hspace{0.4cm} 
When dealing with image submissions, our initial step involves attempting to ascertain the sender's identity. However, not all image submissions include sender information, as it is occasionally cropped out in the submission process. Sender information can manifest in various forms, such as Alphanumeric Sender IDs, standard phone numbers, short codes, custom names or labels, and email addresses. An Alphanumeric Sender ID is typically reserved for one-way messaging from brands and does not allow replies. Standard phone numbers can consist of 7 to 15 digits, often including optional parentheses, hyphens, or spaces. Short codes in the US are typically 5 or 6 digits in length, though different countries might adopt varying lengths. An email address comprises an email prefix and domain, separated by an '@' sign. Despite their varied formats, these types of sender information commonly appear centered in the top section of the image. To pinpoint this information, we utilize the bounding boxes provided by OCR, narrowing our focus to the top 25\% of the message to identify lines potentially containing sender details. Employing a series of regular expressions, we aim to match the text at the top center of the SMS message to identify the sender. However, a complication arises when sender information exceeds the display window, resulting in an abbreviation with '...' to signify the inability to fit within the app's window. Consequently, in some messages, crucial patterns for identifying the sender may be unavailable.

\subsection{Message Attribute Parsing}
\hspace{0.4cm} 
Our technique for parsing the message attributes involved three key steps. Initially, we identify the main body of the text to separate it from any peripheral information or artifacts that might be inherent to default SMS app messaging. Peripheral information may encompass app-generated warnings or notifications related to the message itself. Once the main body of the text is isolated, we proceed to parse it using regular expressions to locate any URLs contained within. In instances where multiple URLs are present, we select one for analysis based on criteria such as URL length, inclusion of paths or query contents. Our aim when dealing with multiple URLs is to prioritize the URL with the most specificity and potential relevance to the message. Subsequently, utilizing the parsed URL and main content from the message, we identify any brands mentioned within the message. These brands could be present in the sender information, content body, or within the URL or redirected URL of the message. Within the dataset, we incorporate a field containing the complete, unfiltered submission text, serving as a reference for understanding the state before and after the parsing process.

\subsection{VirusTotal \& WHOIS Information}
\hspace{0.4cm} 
Following the message parsing, if a URL is present, we conduct a VirusTotal scan and WHOIS request lookup on the URL to acquire historical information about the message at the time of submission. In instances where the message contains a URL shortening service, our WHOIS request is directed to the final URL destination rather than the shortened link.

VirusTotal~\cite{virustotal} generates results using the full URL link extracted from the uploaded image accompanying the message. If the message hasn't undergone prior scanning, VirusTotal initiates a scan on the link. The scan results encompass detection findings from over 70 security vendors, categorizing any detected malicious, malware, or phishing content in the URL. VirusTotal is a crucial element of smishing detection tasks, as it offers a wealth of detection findings. Additionally, by including the scores, we show the room for improvement among detection algorithms, as many message's smishing links are undetected by any of the 70 security vendors. The dataset stores the total detected problems and their respective scores for each category related to the first link in the message. Additionally, we include the count of vendors flagging the message as suspicious, though this doesn't contribute to the detected total.

Subsequently, we retrieve the WHOIS results from the domains of the scanned URLs. While these results offer an extensive array of domain background details, our dataset selectively incorporates the domain registrar, creation time, and last update time. The registrar holds significance, as certain registrars are more commonly abused for hosting malicious websites than others. Furthermore, the creation and last update times serve as indicators of suspicious activity; phishing attacks frequently employ newly created domains just before initiating a campaign. Together, these elements provide important contextual information about the smishing attack at the time it was occurring. 

\subsection{Data Quality}
\hspace{0.4cm} 
To ensure data quality and mitigate the skew of certain messages, our initial step involved removing duplicate entries from the dataset. We established our removal criteria based on identical unfiltered text content extracted from the messages. However, messages with the same main text body but from different sender IDs were preserved in the dataset. Moreover, each message includes a timestamp denoting the moment it was obtained, serving as a reference point for its submission history. Subsequently, we conducted a process to determine the final URL destination for each URL contained within the messages. This involved our efforts to unveil any URL shortening or redirect paths. Lastly, individual message analysis was conducted to ascertain their classification as smishing attacks. This determination relied on a combination of factors, including VirusTotal scores, identification of common smishing tactics, and an analysis encompassing sender information, message content, and domain history.

%% file: results.tex
\section{Message Analysis}
\label{sec:results}
\hspace{0.4cm} Here we breakdown the types of messages collected through smishtank.com. Next, we provide an overview and visualization of the categories of messages and their URLs in the following section. With these steps, we can better understand and interpret the messages gathered from data collection.

\begin{figure}[H]
    \centering
    \includegraphics[width=.4\textwidth]{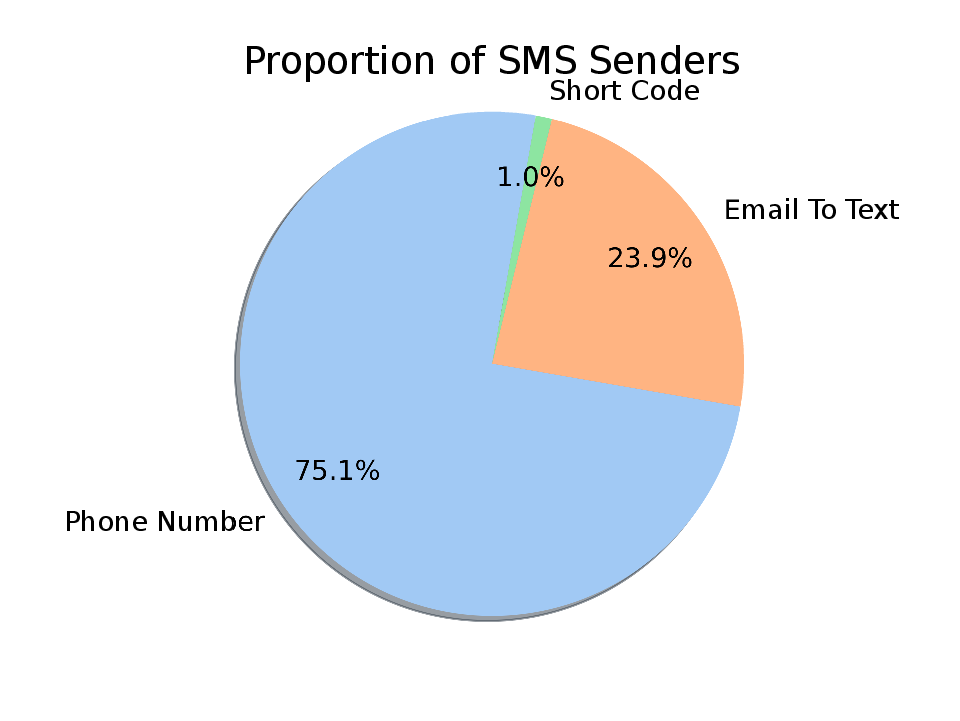}
    \caption{SMS Sender proportions by type.}
    \label{fig:senders}
\end{figure}

\subsection{Sender Information in Messages}
\hspace{0.4cm} 
The sender of an SMS message holds significance in identifying suspicious or fraudulent messages. Typically, large businesses utilize short codes for texting, which are inherently more challenging to spoof compared to long codes. Hence, detecting messages claiming to be from reputable companies but sent from a 10-digit code can be pivotal in flagging potential smishing attempts. Moreover, a considerable portion of messages originated from Email-to-text conversions. In this scenario, an email is transformed and transmitted as an SMS message, with the sender ID displaying the email address. Figure~\ref{fig:senders} illustrates the distribution of each sender type within the SMS messages.

\begin{figure}[h]
    \centering
    \includegraphics[width=.4\textwidth]{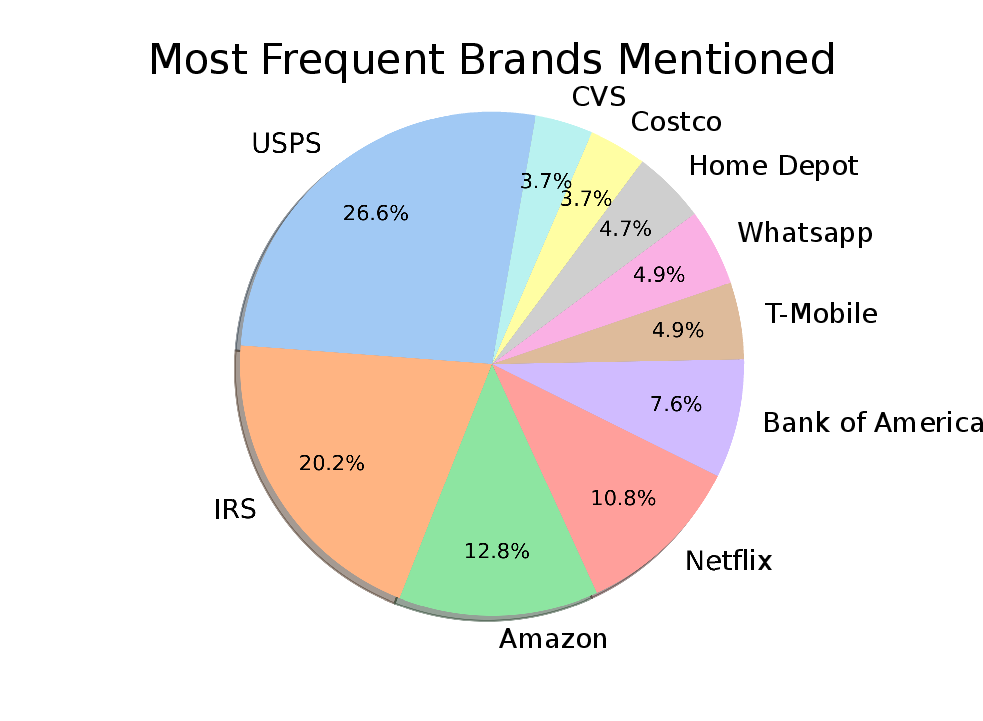}
    \caption{10 most frequent imitated brands in dataset.}
    \label{fig:brands}
\end{figure}

\subsection{Brand Information in Messages}
\hspace{0.4cm} 
In smishing messages, brands are frequently impersonated to deceive users into believing that the messages originate from reputable entities. As a feature within our dataset, we incorporate information on the brands referenced in these messages. Figure~\ref{fig:brands} illustrates the distribution of the most prevalent brands. Notably, the most commonly impersonated brands were associated with product delivery services. These messages often mention a supposed failure in delivering a package and request personal information to proceed with the delivery.

\begin{figure}[H]
    \centering
    \includegraphics[width=.99\linewidth]{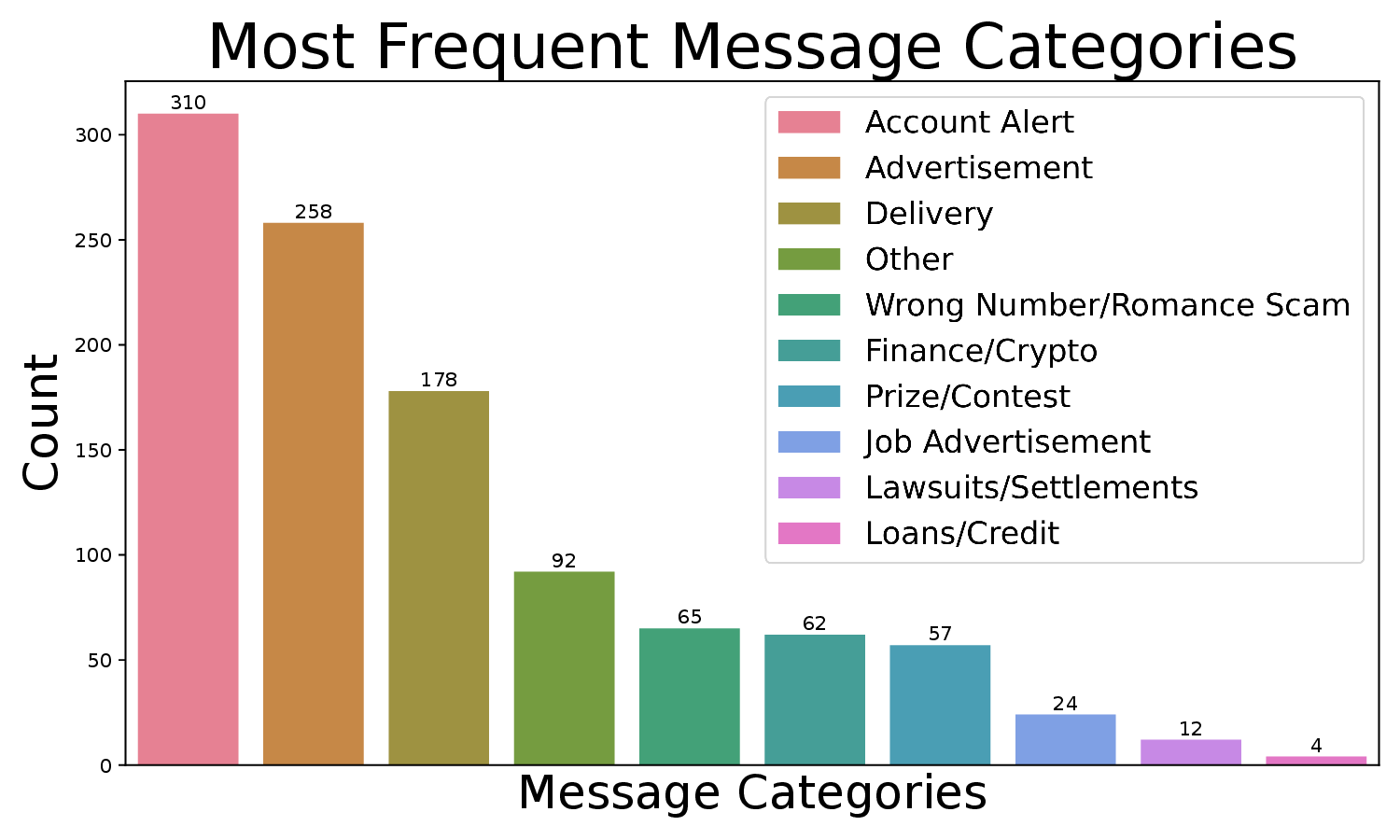}
    \caption{Message Categories.}
    \label{fig:msgcategory}
\end{figure}

\subsection{Message Categorization}
\hspace{0.4cm}
We categorized each message into one of 10 distinct categories, and a comprehensive breakdown of these categories is available in Figure~\ref{fig:msgcategory}. These categories were chosen based on the content of the messages, and the categorization labels were chosen based on previous SMS spam labeling research efforts~\cite{cluesintwitter}. This breakdown illustrates the diverse range of messages submitted to the smishtank.com website. Our collection revealed that the most frequently encountered messages through our platform were advertisements and account alerts.

\begin{figure}[H]
    \centering
    \includegraphics[width=.99\linewidth]{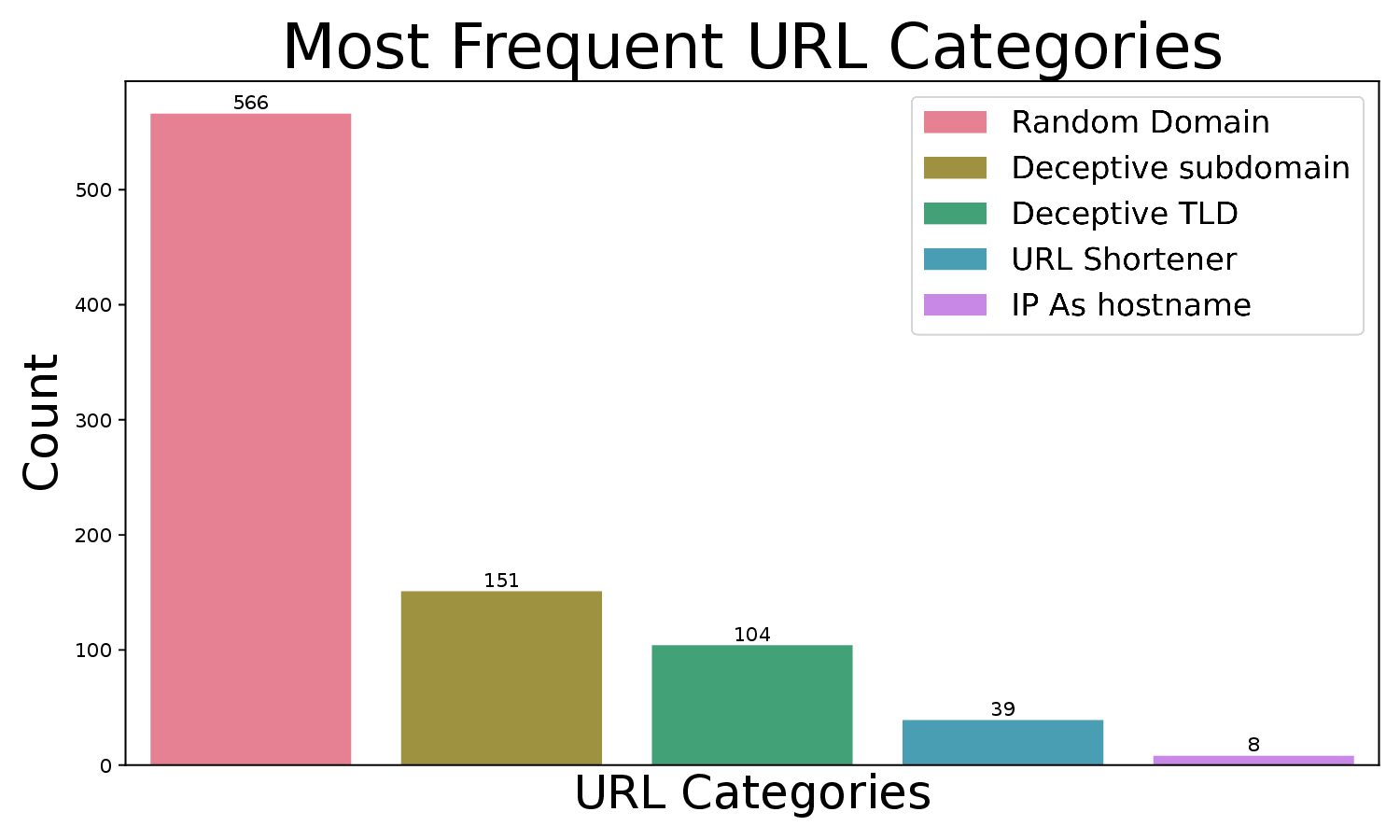}
    \caption{URL Categories}
    \label{fig:urltypes}
\end{figure}

\subsection{URL Categorization}
\hspace{0.4cm}
We classified URLs into five distinct types based on methodologies employed in prior research~\cite{Oest2018}, and the outcomes of this categorization are depicted in Figure~\ref{fig:urltypes}. A Random Domain refers to a URL featuring a domain devoid of any relevant words or terms related to the message content. A deceptive subdomain or Top Level Domain (TLD) involves domains using words or phrases to mislead users into believing it represents an authentic website, with tactics such as Combosquatting and Typosquatting commonly employed. Additionally, some URLs contain the IP address as the domain's hostname. In our analysis, we observed that the majority of messages predominantly utilized random domains that lacked relevance to the message content.

\begin{figure}[H]
        \centering
        \includegraphics[width=.99\linewidth]{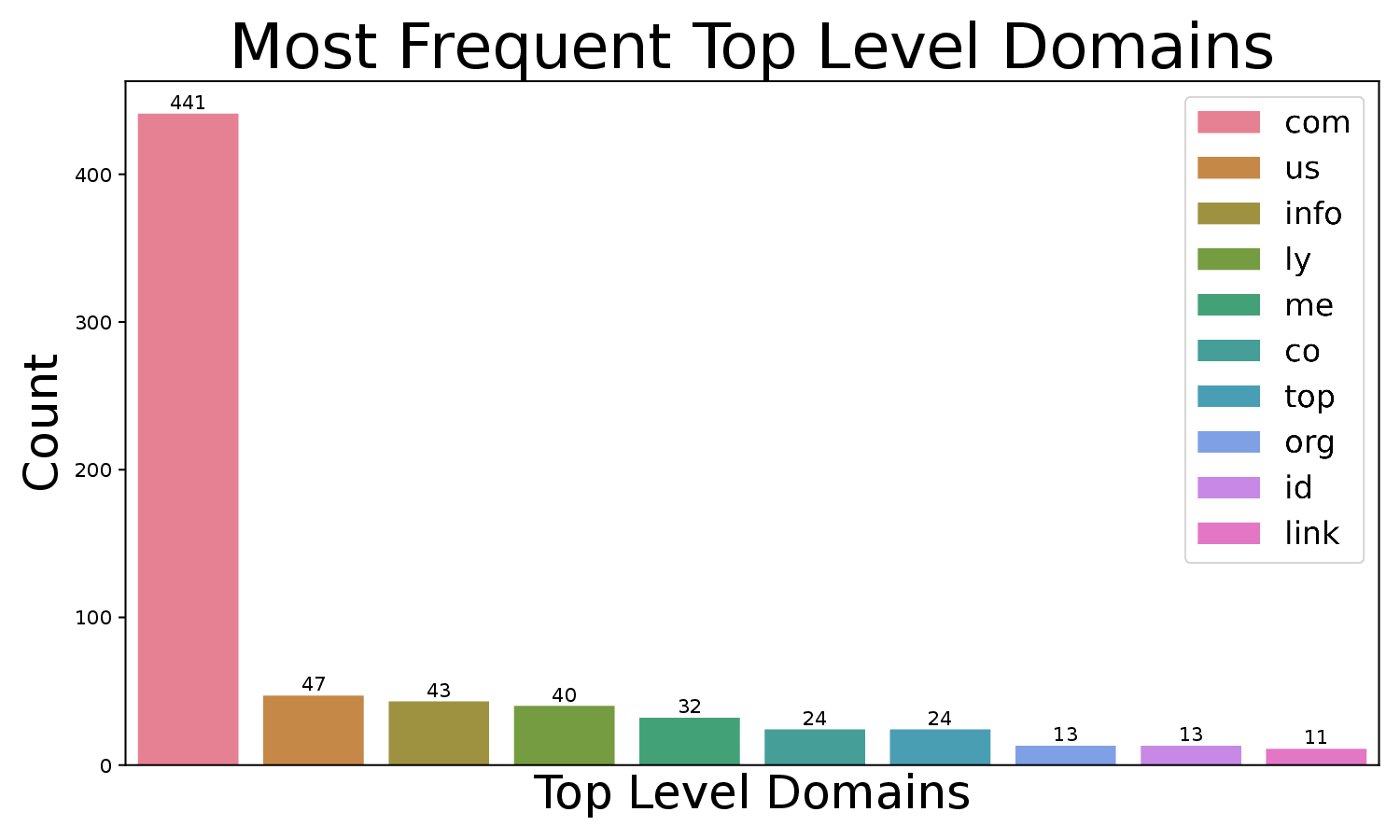}
        \caption{ Top Level Domains}
        \label{fig:squatting}
\end{figure}

\subsection{Top Level Domains}
\hspace{0.4cm} 
The TLD holds significant importance as an indicator for a suspicious website~\cite{Christou_2020}. These domains represent the highest level in the domain hierarchy and appear following the final dot in a website's domain. Phishing domains frequently imitate authentic ones by employing similar names with different TLDs. In Figure~\ref{fig:squatting}, we showcase a list of the ten most commonly encountered TLDs within our dataset.

\begin{table}[H]
\smaller
\begin{tabular}{lcccc}
\hline
Scores & \textgreater{}=1 & \textgreater{}=5 & \textgreater{}=10 & \textgreater{}=15 \\ \hline
Detected & 486 & 203 & 125 & 93 \\
Phishing & 371 & 158 & 95 & 23 \\ \hline
\end{tabular}
\caption{VirusTotal Scores }
\label{tab:vt-results}
\end{table}

\subsection{VirusTotal Scores}
\hspace{0.4cm} 
We utilized the VirusTotal service~\cite{virustotal} to analyze the URLs contained within the messages. To conduct these checks, we integrated the VirusTotal Academic API into our submission process, enabling real-time analysis of messages upon upload. This API performs thorough checks on provided URLs using over 70 antivirus scanners, identifying instances of malware, malicious content, or phishing attempts. Based on these outcomes, we tabulated the total number of detections and categorized them by their respective threat categories. The comprehensive results of our dataset analysis can be found in Table~\ref{tab:vt-results}.

\subsection{Domain Registrars}
\begin{figure}[H]
    \centering
    \includegraphics[width=.99\linewidth]{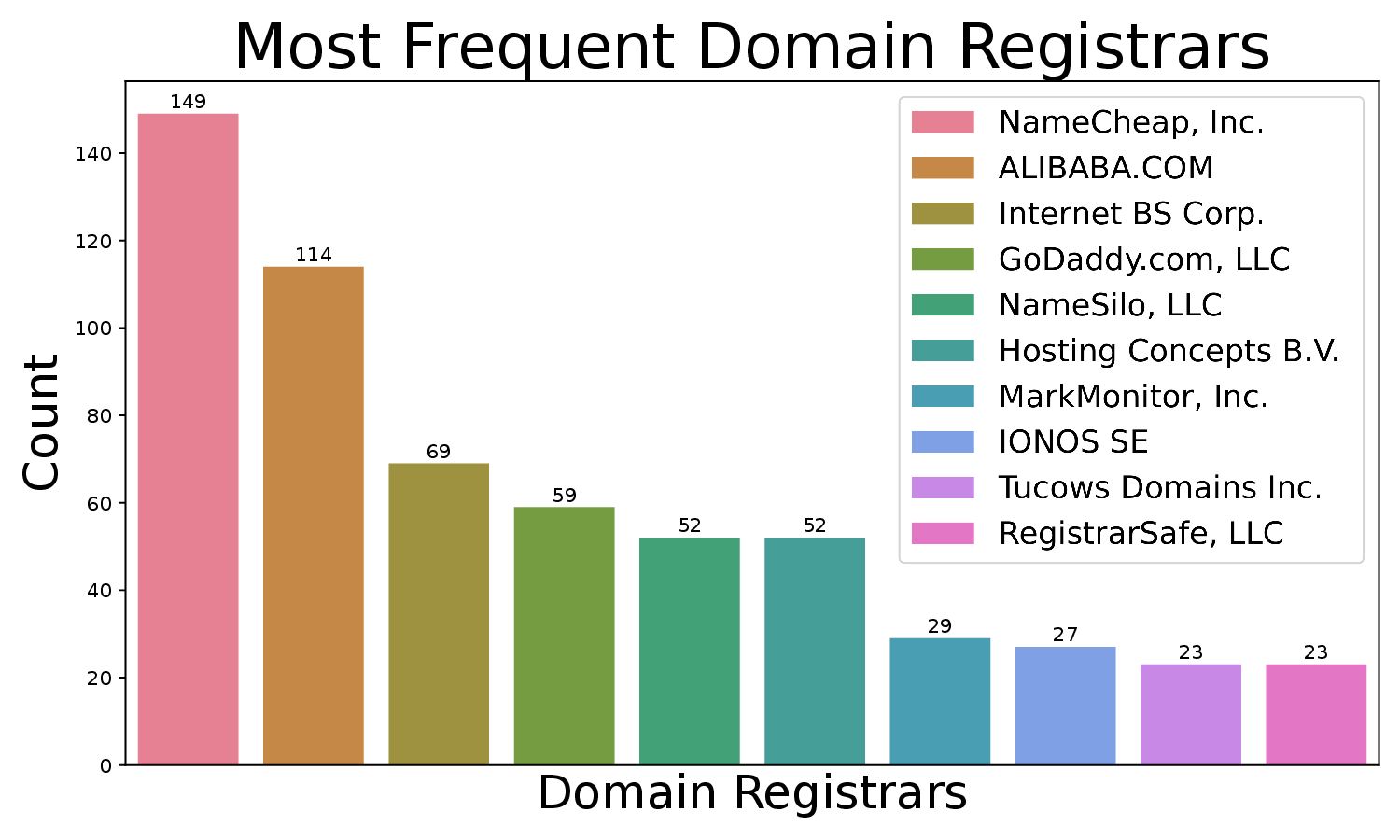}
    \caption{ Domain Registrars}
    \label{fig:domainregisrars}
\end{figure}

\hspace{0.4cm}
We acquired domain histories from the URLs within our dataset by scrutinizing WHOIS data. This analysis allows us to discern patterns in domains utilized within smishing messages. It encompasses trends in domain registrars and details regarding the recency of website creation or modifications in relation to the message reception time. This serves as an indicator of message freshness, as phishing websites are often swiftly erected for campaign distribution and subsequently dismantled. Figure~\ref{fig:domainregisrars} illustrates the top 20 most frequently encountered domain registrars within our dataset.

%% file: discussion.tex
\section{Dataset Distribution}
\label{sec:discussion}

\hspace{0.4cm} The full dataset has been made publicly available through this link~\url{https://smishtank.com/dataset}. This dataset contains each feature from the message parsing process, its VirusTotal information and WHOIS information gathered from the URLs, and categorization of the messages and URLs.

\section{Applications \& Future Work}
\hspace{0.3cm}
The primary aim of this dataset is to offer researchers and developers access to available instances of smishing messages. We perform the analysis of these messasges at the time of their collection to paint a picture of the messages as they are first seen. Researchers can incorporate these messages into their datasets, aiding in smishing detection and other research endeavors. As the collection on smishtank.com continues to grow, the need for larger publicly accessible datasets becomes imperative.

One avenue for future exploration involves analyzing the seasonal or event-based nature of phishing messages. For instance, we observed a surge in IRS-related messages coinciding with the November 2023 extended tax deadline. Identifying smishing campaigns linked to current events and the timing of message receptions could yield valuable insights into phishers' tactics, potentially contributing to the prevention of future campaigns.

%% file: conclusion.tex
\section{Conclusion}
\label{sec:conclusion}

\hspace{0.4cm} 
As smishing attacks evolve, the research community requires an increasing number of smishing data to effectively counter these social engineering attacks. Our efforts have involved collecting, processing, and categorizing 1062 phishing messages, aiming to empower developers and researchers in enhancing detection methods and advancing research on smishing campaigns. To ensure the timeliness of these messages, we've incorporated WHOIS and VirusTotal information, offering insights into their history at the point of collection. We anticipate that this work will lay a crucial foundation for future smishing research.

%% file: acknowledgement.tex
\section{Acknowledgement}
\hspace{0.4cm} The authors thank all participants who submitted smishing samples to \url{https://smishtank.com}, thus enhancing our smishing dataset. We thank all the CODASPY\`24 reviewers for their thoughtful feedback and suggestions.

%% file: main.bbl
\begin{thebibliography}{10}

\bibitem{doi:10.1080/19393555.2015.1078017}
Adesina S.~Sodiya Adebukola S.~Onashoga, Olusola O. Abayomi-Alli and David~A. Ojo.
\newblock An adaptive and collaborative server-side sms spam filtering scheme using artificial immune system.
\newblock {\em Information Security Journal: A Global Perspective}, 24(4-6):133--145, 2015.

\bibitem{apwg}
Apwg | unifying the global response to cybercrime.
\newblock https://apwg.org/.

\bibitem{8965429}
Caner Balim and Efnan~Sora Gunal.
\newblock Automatic detection of smishing attacks by machine learning methods.
\newblock In {\em 2019 1st International Informatics and Software Engineering Conference (UBMYK)}, pages 1--3, 2019.

\bibitem{DatingPhish}
Vincent Drury, Luisa Lux, and Ulrike Meyer.
\newblock Dating phish: An analysis of the life cycles of phishing attacks and campaigns.
\newblock ARES '22, New York, NY, USA, 2022. Association for Computing Machinery.

\bibitem{FCCRobotext}
FCC.
\newblock Robotext scams on the rise, 2022.
\newblock Accessed on 12 10, 2023.

\bibitem{FCCRulesChange}
FCC.
\newblock Fcc adopts its first rules focused on scam texting, 2023.
\newblock Accessed on 12 10, 2023.

\bibitem{10139070}
Shaghayegh Hosseinpour and Hadi Shakibian.
\newblock An ensemble learning approach for sms spam detection.
\newblock In {\em 2023 9th International Conference on Web Research (ICWR)}, pages 125--128, 2023.

\bibitem{irswarn}
IR-2022-167.
\newblock Irs reports significant increase in texting scams; warns taxpayers to remain vigilant, 2022.
\newblock Accessed: Dec 29, 2023.

\bibitem{Mishra2021DSmishSMSAST}
Sandhya Mishra and Devpriya Soni.
\newblock Dsmishsms-a system to detect smishing sms.
\newblock {\em Neural Computing \& Applications}, pages 1 -- 18, 2021.

\bibitem{nahapetyansms}
Aleksandr Nahapetyan, Sathvik Prasad, Kevin Childs, Adam Oest, Yeganeh Ladwig, Alexandros Kapravelos, and Bradley Reaves.
\newblock On sms phishing tactics and infrastructure.
\newblock In {\em 2024 IEEE Symposium on Security and Privacy (SP)}, pages 169--169. IEEE Computer Society, 2024.

\bibitem{insidephishersmind}
Adam Oest, Yeganeh Safei, Adam Doupé, Gail-Joon Ahn, Brad Wardman, and Gary Warner.
\newblock Inside a phisher's mind: Understanding the anti-phishing ecosystem through phishing kit analysis.
\newblock In {\em 2018 APWG Symposium on Electronic Crime Research (eCrime)}, pages 1--12, 2018.

\bibitem{Oest2018}
Adam Oest, Yeganeh Safei, Adam Doupé, Gail-Joon Ahn, Brad Wardman, and Gary Warner.
\newblock Inside a phisher's mind: Understanding the anti-phishing ecosystem through phishing kit analysis.
\newblock In {\em 2018 APWG Symposium on Electronic Crime Research (eCrime)}, pages 1--12, 2018.

\bibitem{phishtank}
{Join the fight against phishing} phishtank.
\newblock \url{https://phishtank.org/}.

\bibitem{proofpointsotf}
proofpoint.
\newblock 2023 state of the phish report, 2023.
\newblock Accessed on 12 10, 2023.

\bibitem{bethephisher}
Florian Quinkert, Martin Degeling, Jim Blythe, and Thorsten Holz.
\newblock Be the phisher -- understanding users' perception of malicious domains.
\newblock In {\em Proceedings of the 15th ACM Asia Conference on Computer and Communications Security}, ASIA CCS '20, page 263–276, New York, NY, USA, 2020. Association for Computing Machinery.

\bibitem{Raman2023}
Md~Lutfor Rahman, Daniel Timko, Hamid Wali, and Ajaya Neupane.
\newblock Users really do respond to smishing.
\newblock In {\em Proceedings of the Thirteenth ACM Conference on Data and Application Security and Privacy}, CODASPY '23, page 49–60, New York, NY, USA, 2023. Association for Computing Machinery.

\bibitem{reaves2018}
Bradley Reaves, Luis Vargas, Nolen Scaife, Dave Tian, Logan Blue, Patrick Traynor, and Kevin R.~B. Butler.
\newblock Characterizing the security of the sms ecosystem with public gateways.
\newblock {\em ACM Trans. Priv. Secur.}, 22(1), dec 2018.

\bibitem{Saeki2022}
Ryu Saeki, Leo Kitayama, Jun Koga, Makoto Shimizu, and Kazumasa Oida.
\newblock Smishing strategy dynamics and evolving botnet activities in japan.
\newblock {\em IEEE Access}, 10:114869--114884, 2022.

\bibitem{salman2022empirical}
Muhammad Salman, Muhammad Ikram, and Mohamed~Ali Kaafar.
\newblock An empirical analysis of sms scam detection systems, 2022.

\bibitem{Profiler2022}
Mariya Shmalko, Alsharif Abuadbba, Raj Gaire, Tingmin Wu, Hye-Young Paik, and Surya Nepal.
\newblock Profiler: Distributed model to detect phishing.
\newblock In {\em 2022 IEEE 42nd International Conference on Distributed Computing Systems (ICDCS)}, pages 1336--1337, 2022.

\bibitem{virustotal}
Gaurav Sood.
\newblock {\em virustotal: R Client for the virustotal API}, 2021.
\newblock R package version 0.2.2.

\bibitem{cluesintwitter}
Siyuan Tang, Xianghang Mi, Ying Li, XiaoFeng Wang, and Kai Chen.
\newblock Clues in tweets: Twitter-guided discovery and analysis of sms spam, 2022.

\bibitem{timkounveil23}
Daniel Timko, Daniel~Hernandez Castillo, and Muhammad~Lutfor Rahman.
\newblock Unveiling human factors and message attributes in a smishing study.
\newblock {\em arXiv preprint arXiv:2311.06911}, 2023.

\bibitem{timko2023commercial}
Daniel Timko and Muhammad~Lutfor Rahman.
\newblock Commercial anti-smishing tools and their comparative effectiveness against modern threats.
\newblock In {\em Proceedings of the 16th ACM Conference on Security and Privacy in Wireless and Mobile Networks}, pages 1--12, 2023.

\end{thebibliography}
